\documentstyle[12pt,epsfig,cite]{article}
\title{The decay of plane wave pulses with complex structure in a nonlinear
dissipative medium}
\author{Sergei~N.~Gurbatov, Bengt~O.~Enflo, Galina~V.~Pasmanik} 
\begin{document} \date{} 
\maketitle

\vfill
\begin{center}
\begin{tabular}{ll}

& Radiophysics Dept., University of Nizhny Novgorod \\
& 23, Gagarin Ave., Nizhny Novgorod 603600, RUSSIA \\
& e-mail: gurb@rf.unn.runnet.ru\\
\\
& Department of Mechanics,\\
& Royal Institute of Technology,\\
& S--100 44 Stockholm, SWEDEN\\
& e-mail: benflo@mech.kth.se,\\
& phone: Int+46 8 7907156, fax: Int+46 8 7969850\\

\end{tabular}
\end{center}
\vfill
\pagebreak
\begin{abstract}
Nonlinear plane acoustic waves propagating through a fluid are studied
using Burgers'
equation with finite viscosity. The evolution of a simple N-pulse with
regular and random initial
amplitude and of pulses with monochromatic and noise carrier is considered.
In the latter case the
initial pulses are characterized by two length scales. The length scale of
the modulation function
is much greater than the period or the length scale of the carrier. With
increasing time the initial
pulses are deformed and shocks appear. The finite viscosity leads to a
finite shock width, which does
not depend on the fine structure of the initial pulse and is fully
determined by the shock position
in the zero viscosity limit. The other effect of nonzero viscosity is the
shift of the shock
position from the position at zero viscosity. This shift, as well as the
linear time, at which
the nonlinear stage of evolution changes to the linear stage, depends on
the fine structure of
the initial pulse. It is also shown that the nonlinearity of the medium
leads to generation of a
nonzero mean field from an initial random field with zero mean value. The
relative fluctuation of
the field is investigated both at the nonlinear and the linear stage.
\end{abstract}
\pagebreak

\section{Introduction}
The propagation of finite amplitude sound waves is of fundamental interest
in nonlinear acoustics.
In the simplest model of propagation in fluids these waves are described by
the well-known Burgers' equation (plane waves) \cite{Burgers74,Whitham}
or modifications of Burgers' equation, which are called generalized
Burgers' equations (cylindrical and spherical waves) \cite{RudenkoSoluyan,
Crighton,Sachdev87}.
In studies of nonlinear wave propagation an important problem is to find
the waveform of the
asymptotic wave at long time after the preparation of the initial wave or
at long distance from
the source emitting the wave.
In the first case the asymptotic wave is
called the old-age wave and
is an asymptotic solution of an initial value problem of some wave equation
and in the latter case
the asymptotic wave is a solution of a boundary wave problem.
The asymptotic wave is damped both by
absorption and by shock wave dissipation and the asymptotic wave is
therefore, for plane waves,
described by the linear diffusion equation.
Because the linear diffusion equation describes the
attenuation of high frequency waves, the asymptotic behaviour is determined
by the spectrum of the
waves of low frequencies.
This spectrum is the result of the distortion of the wave at the nonlinear
stage.

The decay of a wave to the old-age waveform is very different for periodic
signals and pulse perturbations.
For the periodic signal the old-age waveform is an
exponentially decaying harmonic wave.
The pulse perturbation has continuous spectrum and the amplitude of
the pulse decays according to a power law.

For sinusoidal and N-wave initial perturbations the old-age solutions have
been studied in several
papers for plane, cylindrical and spherical waves with use of both
analytical and numerical methods
\cite{CrightonScott,Scott,Enflo81,Enflo85,Sachdev86,Sachdev89,Hammerton,
Enflo93,Enflo96,Enflo98}.
The dependence of the amplitude constant of the old-age waveform on
the parameters of the initial perturbations has been found.

The aim of the present paper is to investigate the asymptotic behaviour of
complex pulses, such as modulated or random waves.
Plane waves are studied, which means that we will use the original form
of Burgers' equation \cite{Burgers74}.
Burgers' equation has an exact solution, which is found by reducing
it to the linear diffusion equation by mean of the Hopf-Cole transformation
\cite{Cole,Hopf50}.
Because of the existence of an exact solution it is relatively easy to
find the old-age waveform developing
from simple initial perturbations like periodic signals and N-waves in the
plane wave case.
However, for signals with complex structure it is far from trivial to
find their evolution even for plane waves
\cite{GuCr95,Brander,AngVas98}.
For a regular signal with fractal structure \cite{GuCr95,AngVas98}
an unusual sequence of stages of evolution may appear: the nonlinear
stage may succeed the linear stage.
At the nonlinear stage this
wave may decrease more slowly than a periodic wave or a simple pulse.

Still more complicated behaviour is shown by a random signal.
The solution of Burgers' equation
with random initial conditions is often called Burgers turbulence.
Numerous papers are devoted to
investigations of this problem \cite{Esipov,AMS94,KPZ,MSW95,GSAFT97,
Molchan97}.
In the case of vanishing viscosity the
continuous random initial field is transformed into a sequence of straight
lines with some slope and with random locations of the shocks separating them.
Due to the merging of the shocks the
internal scale of the turbulence increases and the random wave decreases
more slowly than the periodic signal.
The decay rate depends on the behaviour of the initial energy spectrum of
low frequency \cite{GMS91,GSAFT97} and is also sensitive to the
statistics of the potential of the initial velocity field
\cite{Esipov,GSAFT97,Molchan97}.
The asymptotic behaviour of the field in the case of finite viscosity also
depends strongly on the statistical properties of the initial field.
In the case where there are
large scale components in the initial spectrum the nonlinear stage never
transforms to the linear regime of evolution \cite{GMS91}.
In the opposite case the final evolution depends on the tail of the
initial probability  distribution \cite{AMS94}.

In the present paper we consider the evolution of complex pulses which are
characterized by two
scales: $l_{0}$ - the inner scale of the carrier, $L_{0}$ - the scale of
the modulation, and the condition $L_{0} >> l_{0}$.
For such signals the generation of a low-frequency component or a
non-zero mean field takes place.
The case of vanishing viscosity has been investigated \cite{Gurbatov82,
GEP99}.
It has been shown that, after multiple merging of the shocks, the inner
structure disappears and finally a finite pulse ends up with an N-wave.
This N-wave is fully described by the positions of its zero
and shocks, which are determined by the potential $\psi_{0}(x)$ of the
initial velocity $v_{0}(x)$: $v_{0}'(x)=-\psi_{0}(x)$.
It has also been shown that, for a pulse with random carrier, the
fluctuations of the positions of the zero and the shocks are relatively
small for $l_{0}<<L_{0}$.

In the present paper we will consider the influence of finite
viscosity on the nonlinear stage of the evolution of the pulse and
investigate its old-age behaviour.
The paper is organized as follows.

In section 2, on the base of the Hopf-Cole solution, the asymptotic
behaviour of the pulse at the
nonlinear and the old-age stages is analyzed in general terms without
assuming the detailed structure of the initial pulse.

In section 3  the evolution of simple N-waves with regular and random
amplitude is discussed.

Section 4 deals with pulses with monochromatic carrier and section 5 with
pulses with noise carrier.
It is shown that the parameters of the asymptotic waveform depend
weakly on the fine
structure of the initial pulse, but that the old-age behaviour is very
sensitive to the properties of the carrier.

\section{Solution of Burgers' equation and its large time asymptotics}

Our starting-point is the Burgers' equation
\begin{equation}
\frac{\partial v}{\partial t} + v \frac{\partial v}{\partial x} =
\nu\frac{\partial^2 v}{\partial x^2} \;\;,
\label{beq}
\end{equation}
which governs the propagation of plane acoustic waves in nonlinear
dissipative media.
Here $v$ is the velocity of fluid and $\nu$ is the fluid viscosity.
Burgers' equation (\ref{beq}) has the Hopf-Cole solution \cite{Cole,Hopf50}
\begin{equation}
v(x,t) = -2\nu\frac{U_x}{U} \;\;,
\label{hc}
\end{equation}
where $U(x,t)$ is the solution for the linear diffusion equation
\begin{equation}
U(x,t) = \int\limits^{+\infty}_{-\infty} U_0(y)G(x,y,t)dy \;\;.
\label{sold}
\end{equation}
Here $G(x,y,t)$ is the Green function of the linear diffusion equation
\begin{equation}
G(x,y,t) = \frac{1}{\sqrt{4\pi\nu t}}\exp\left[-\frac{(x-y)^2}
{4\nu t}\right]\;\;,
\label{green}
\end{equation}
and $U_0(x)$ is the initial condition for this equation
\begin{equation}
U(x,0)=U_0(x)=\exp\left[\frac{\psi_0(x)}{2\nu}\right]\;\;,
\label{incon}
\end{equation}
which corresponds to the initial condition
\begin{equation}
v(x,0)=v_0(x)=-\frac{d\psi_0(x)}{dx} \;\;.
\label{invel}
\end{equation}

The main goal of the present paper is the investigation of the old-age
behaviour of the localized pulses.
But first we discuss shortly the asymptotic evolution of nonlocalized waves.
Consider a group of perturbation with the bounded initial potential
$|\psi_0(x)|<\infty$ assuming that $\psi_0(x)$ is a periodic signal
with a period $L_0$ or homogeneous noise with rather fast decreasing
probability distribution of the potential $\psi_0$.
For such a perturbation in $U(x,t)$ (\ref{sold}) we separate a constant
component $\bar{U}$:
\begin{equation}
U(x,t)=\bar{U}+\tilde{U}(x,t) \;\;.
\label{inu}
\end{equation}
Inserting (\ref{inu}) into (\ref{sold}) we see that $\bar U$ does not depend
on time, since the $y$-integral of $G(x,y,t)$ is unity.
Here $\tilde{U}(x,t)$ is a field with zero mean value (on the period or
statistically for noise).
As times goes on, the viscous dissipation and oscillation (inhomogenity)
smoothing causes the amplitude (variance) of the field $\tilde{U}(x,t)$
to become less.
At times when it amounts to  $|\tilde{U}|\ll \bar{U}$ the solution
(\ref{hc}) is equal to
\begin{equation}
v(x,t)=-2\nu\frac{\tilde{U}_x(x,t)}{\bar U} \;\;.
\label{uhc}
\end{equation}
As $\tilde{U}(x,t)$ satisfies the linear diffusion equation, then $v(x,t)$
also at these times fulfils the linear equation.
This testifies precisely to the fact that the evolution of the velocity
field enters the linear stage.
The accumulated nonlinear effects are described in this solution by the
nonlinear integral relation between the initial velocity field $v_0(x)$
and the fields $\tilde U (x,0)$, $\bar U$ (\ref{incon},\ref{invel}), and are
characterized by the value $|\triangle\psi_0|/\nu\sim Re_0$.

Here $\triangle\psi_0$ is the characteristic change in amplitude of
 $\psi_0$, and $Re_0$ is the initial Reynolds number.

 From (\ref{uhc}) it is easy to get the well known result, that for
 $Re_0\gg 1$ the initial harmonic waves asymptotically has also
 harmonics form, but with the amplitude not depending on the
 initial amplitude \cite{Whitham,RudenkoSoluyan}.

 At large initial Reynolds number the homogeneous Gaussian field $v_0(x)$
 at the nonlinear stage transforms into series of sawtooth waves with strong
 nongaussian statistical properties \cite{GMS91, MSW95}.
 Nevertheless at very large time, when the relation (\ref{uhc}) is valid,
 the distribution of the random field $v(x,t)$ with statistically homogeneous
 initial potential $\psi_0(x)$ converges weakly to the distribution of the
 homogeneous Gaussian random field with zero mean value, and universal
 covariance function \cite{AMS94}.
 But the amplitude of this function is nonlinearly related to the initial
 covariance function of the field $v_0(x)$ and increases proportionally
 $\exp(Re^2_0)$ with increasing of initial Reynolds number $Re_0$
 \cite{GMS91,AMS94}.

 Let us consider the opposite situation when the initial perturbation is
 localized in some region.
 We assume also that initial potential $\psi_0(x)$ vanishes very fast for
 $x\to\infty$, so we have
 \begin{equation}
 \psi_0(x)=0\;\;,\;\;\;\; |x|>L_* \;\;,
 \label{inpot}
 \end{equation}
 where $L_*$ can be considered as the length of initial pulse.
 The condition (\ref{inpot}) is identical to the assumption that
 the integral over the initial pulse is zero.

 It should be stressed that if the integral over the
 initial pulse is $\triangle\psi_0\neq 0$, then the initial
 perturbation transforms to the unipolar pulse with area $\triangle\psi_0$
 asymptotically.
 For $Re_0=\triangle\psi_0/\nu\gg 1$ the form of this pulse is close to
 the triangular with the width of shock $\delta(t)\sim\nu t/x_s$ much
 smaller than the length of the pulse $x_s=\sqrt{2|\triangle\psi_0|t}$.
 For such a pulse $\delta(t)/x_s\sim Re_0^{-1}=const$ and does not
 depend on time.
 Thus the nonlinear stage of the evolution prevails for all times.

 Let us now consider the time asymptotic behaviour of initially localized
 pulse which satisfies (\ref{inpot}).
 If we rewrite (\ref{sold}) as
 \begin{equation}
 U(x,t)=1+\int\limits_{-\infty}^{+\infty}[U_0(y)-1]G(x,y,t)dy \;\;,
 \label{resold}
 \end{equation}
 then we can see that because of (\ref{inpot}) the integrand at right
 hand side of (\ref{resold}) is zero outside the region $|y|<L_*$.
 It is seen from (\ref{green}) that the length scale $l_{dif}(t)$ of
 Green's function $G(x,y,t)$ is
 \begin{equation}
 l_{dif}=\sqrt{2\nu t} \;\;.
 \label{ldif}
 \end{equation}
 If the length scale of $U_0(y)$ is called $L_U$, and the following
 condition
 \begin{equation}
 l_{dif}(t)\gg L_U
 \label{lL}
 \end{equation}
 is valid, then Green's function can be considered as constant in the
 interval where the integrand at the right hand side of (\ref{resold})
 is significantly different from zero.
 Thus we obtain from (\ref{resold}) using (\ref{incon})
 \begin{equation}
 U(x,t)\approx 1 + G(x,y^*(x,t),t)B \;\;.
 \label{uapprox}
 \end{equation}
 Here $y^*(x,t)$, $|y^*|<L_*$, is the value of $y$, in the neighborhood
 of which the integrand in (\ref{resold}) gives the essential contribution
 in the integral.
 In (\ref{uapprox}) we introduce the notation
 \begin{equation}
 B = \int\limits_{-\infty}^{+\infty}\left[\exp\left(\frac{\psi_0(y)}{2\nu}
 \right)-1\right]dy \;\;.
 \label{b}
 \end{equation}
 We assume $B\neq 0$. From (\ref{sold}),(\ref{uapprox}),(\ref{b}) an
 approximate solution is obtained
 \begin{equation}
 v(x,t)=\frac{x-y^*}{t}\frac{BG(x,y^*,t)}{1+BG(x,y^*,t)} \;\;.
 \label{apprsol}
 \end{equation}
 If the initial pulse $\psi_0$ is centered around $y=0$, and the large scale
 of the field $v(x,t)$ is much greater then $L_*$, then we can put
 $y^*\approx 0$ and obtain
 \begin{equation}
 v(x,t)=\frac{x}{t}\frac{BG(x,0,t)}{1+BG(x,0,t)}
 \label{apsol0}
 \end{equation}
 instead of (\ref{apprsol}).
 For large initial Reynolds number $Re_0=A/2\nu$, where $A>0$ is the
 maximum of $\psi_0(y)$, the constant $B$ (\ref{b}) may be rewritten
 as the product of the maximum of the integrand and some length:
 $L_{eff}$
 \begin{equation}
 B=L_{eff}e^{A/2\nu}\;\;,\;\;\;\;
 Re_0=A/2\nu \;\;,
 \label{BRe}
 \end{equation}

 Using the definition of Green's function (\ref{green}) and (\ref{BRe})
 we can rewrite (\ref{apsol0}) in the form
 \begin{equation}
 v(x,t)=\frac{x}{t}\frac{\exp\left[-\displaystyle\frac{1}{2\nu}\left(
 \frac{x^2}{2t}-A+\nu\ln\frac{4\pi\nu t}{L^2_{eff}}\right)\right]}
 {1+\exp\left[-\displaystyle\frac{1}{2\nu}\left(
 \frac{x^2}{2t}-A+\nu\ln\frac{4\pi\nu t}{L^2_{eff}}\right)\right]}
 \;\;.
 \label{solv}
 \end{equation}
 In the limit of vanishing viscosity ($\nu\to 0$) we get from
 (\ref{solv}) the well known solution for N-wave
 \begin{equation}
 v(x,t) = \left\{
         \begin{array}{lll}
         \displaystyle\frac{x}{t}, & |x|<x_s \\ \\
         0, & |x|>x_s,
         \end{array}
         \right. \;\;,
 \label{wknsol}
 \end{equation}
 where $x_s$ is the position of the shock
 \begin{equation}
 x_s=\sqrt{2At} \;\;.
 \label{xs}
 \end{equation}
 At this stage the form and the energy of the pulse
 \begin{equation}
 E(t)=\int v^2(x,t)dx =\displaystyle\frac{2^{5/2}A^{3/2}}{3t^{1/2}}
 \label{E}
 \end{equation}
 are determined only by the value of maximum of initial potential $A$ and
 does not depend on the fine structure of the pulse.
 This limiting case ($\nu\to 0$) for the pulses with complex inner structure
 was investigated in the paper \cite{GEP99}.

 For large but finite Reynolds number the solution (\ref{solv}),
 completed with a solution valid at $x$ near $x_s$ and exhibiting the
 shock structure, is valid only at finite time.
 For finite $t$ values and sufficiently small $x$ we still have
 \begin{equation}
 BG(x,0,t)\gg 1 \;\;.
 \label{bg}
 \end{equation}
 Then for
 \begin{equation}
 |x|\le x_s = \sqrt{2t\left(A-\nu\ln\displaystyle
 \frac{4\pi\nu t}{L^2_{eff}}\right)}{}\;\;,
 \label{23}
 \end{equation}
 \begin{equation}
 BG(x_s,0,t)=1
 \label{bg1}
 \end{equation}
 we obtain the increasing part (in $x$) of the N-wave solution
 (\ref{wknsol}).
 On the other hand, for $x\to\infty$, $t$ fixed, we
 find that $v(x,t)$ fades away exponentionally.

 For $t\to\infty$, $x$ arbitrary finite, we have
 \begin{equation}
 BG(x,0,t)\ll 1 \;\;,
 \label{bg2}
 \end{equation}
 and the solution (\ref{apsol0}) can be approximated as
 \begin{equation}
 v(x,t)=B\frac{x}{\sqrt{4\pi\nu t^3}}\exp\left[-\frac{x^2}{4\nu t}\right].
 \label{vapprox}
 \end{equation}
 Using (\ref{green}) we can rewrite the condition of the validity of the
 solution (\ref{vapprox}) as
 \begin{equation}
 \sqrt{4\pi\nu t}\gg B \;\;\ , \;\;\;\;
 \sqrt{\displaystyle\frac{4\pi\nu t}{L_{eff}^2}}\gg e^{A/2\nu} \;\;.
 \label{recond}
 \end{equation}
 Because (\ref{bg1}) gives a solution of the linear diffusion equation,
 the condition (\ref{recond}) defines the linear stage of the pulse evolution.
 On this stage the pulse energy  is
 \begin{equation}
 E(t)=B^2\sqrt{\displaystyle\frac{\nu}{8\pi t^3}} \;\;.
 \label{30}
 \end{equation}

 Thus at the linear stage of evolution of the pulse has a universal form
 (\ref{vapprox}) and is determined only by the constant $B$, which is defined by
 the initial perturbation by the relation (\ref{b}).
 The case $B=0$ can be understood by comparing the definition of $B$
 according to the equations (14)
 and (5) with the formulas (30) - (33) in ref. [19]. By this comparison it
 is clearthat the case
 $B=0$ corresponds to the absence of the Fourier component with $n=1$ in the
 terminology of ref. [19].
 The Fourier component with $n=0$ is absent here already for $B\neq 0$. The
 case $B=0$ excludes the
 N-wave solution (19).

 Below we will consider three cases of initial perturbation, assuming that
 the initial potential may be written in the form
 \begin{equation}
 \psi_0(x)=M(x)F(x) \;\;,
 \label{31}
 \end{equation}
 where $M(x)$ has the scale $L_0\simeq L_*$.
 First we consider the simplest case when $F$ is constant, either
 deterministic or random value.
 Here we discuss the main properties
 of the wellknown solution (see ref.  \cite{SJN94}) for the N-wave.
 The cases of monochromatic and noise carrier will be considered in Sections
 4 and 5.
 In these cases we assume that the scale $l_0$ of the carrier
 $F(x)$ satisfies the condition $l_0\ll L_0$ and then
 \begin{equation}
 v(x,0)=v_0(x)\simeq M(x)f(x)\;\;, \;\;\;\; f(x)=-F'(x) \;\;.
 \label{32}
 \end{equation}

 \section{The decay of a simple pulse with random initial amplitude}

 \subsection{The evolution of a regular N-wave}

 We discuss the evolution of a one scale pulse, whose potential has the
 structure in (\ref{31}):
 \begin{equation}
 \psi_{0}(x) =M(x)A.
 \label{33}
 \end{equation}
 Here $A$ is a constant and $M(x)$ is a dimensionless function with the
 scale $L_{0}$:
 \begin{equation}
 M(x) \approx 1-\frac{x^2}{2L_{0}^2}+... \;\;.
 \label{34}
 \end{equation}
 In particular we will consider the cases where the initial perturbation
 (\ref{33}), (\ref{34}) is exact in the interval $|x| < \sqrt{2}L_{0}$ and
 $\psi_{0}(x) \equiv 0$ outside this interval.
 We first assume $A>0$ and find from (\ref{invel}) that the initial velocity
 pulse is an N-wave:
 \begin{eqnarray}
 v_{0}(x) = \frac{xA}{L_{0}}, \, |x|<\sqrt{2}L_{0},\nonumber\\
 v_{0}(x) = 0, \, |x|>\sqrt{2}L_{0}.
 \label{35}
 \end{eqnarray}
 From (\ref{incon}) we then obtain
 \begin{eqnarray}
 U_{0} (x) = \exp\left(\frac{\psi_{0}(x)}{2\nu}\right) =
 \exp\left[\frac{A}{2\nu}\left(1-\frac{x^2}{2L_{0}^2}\right)\right], \,
 |x|<\sqrt{2}L_{0} \;\; ;\nonumber\\
 U_{0}(y) =1, \, x>\sqrt{2}L_{0}.
 \label{36}
 \end{eqnarray}
 From (\ref{36}) we find that for this special case the length scale
 $L_{U}$ of $U_{0}(y)$, first
 introduced in (\ref{ldif}), is (the square root of the inverted coefficient
 of $x^2$ in (\ref{36}))
 \begin{equation}
 L_{U}\simeq L_{0}(\frac{2\nu}{A})^{\frac{1}{2}} = \frac{L_{0}}{{\rm
 Re}_{0}^{\frac{1}{2}}},
 \;\;\;\; {\rm Re}_{0} \equiv \frac{A}{2\nu}.
 \label{37}
 \end{equation}

 The calculation of $U(x,t)$, from which we obtain $v(x,t)$ by use of
 (\ref{hc}), can now be done using
 (\ref{resold}) with insertion of (\ref{36}) and (\ref{green}):
 \begin{equation}
 \begin{array}{l}
 U(x,t) = 1+\displaystyle\frac{1}{\sqrt{4\pi\nu t}}\int_{-\sqrt{2}L_{0}}
 ^{\sqrt{2}L_{0}}
 \displaystyle\left(\exp\left[\frac{A}{2\nu}\left(1-\frac{y^2}{2L_{0}^2}\right)
 \right]-1\right)\,\exp\left[-\frac{(x-y)^2}{4\nu t}\right] dy=
 \\
 =1+\displaystyle\frac 12
 \frac{1}{\left(1+\displaystyle\frac{At}{L_{0}^2}\right)^{1/2}}\,
 \exp\left[\frac{A}{2\nu}- \displaystyle\frac{A}{4\nu
 L_{0}^2}\displaystyle\frac{x^2}{1+\displaystyle\frac{At}
 {L_{0}^2}}\right]\cdot
 \\
 \left\{ {\rm erf}\left[\left(\sqrt{2}L_{0}-\displaystyle\frac{x}{1+
 \displaystyle\frac{At}{L_{0}^2}}\right)\sqrt{\displaystyle\frac{1}{4\nu t}
 \left(1+\displaystyle\frac{At}{L_{0}^2}\right)}\right]+
 {\rm erf}\left[\left(
 \sqrt{2}L_{0}+\displaystyle\frac{x}{1+\displaystyle\frac{At}{L_{0}^2}}\right)
 \sqrt{\displaystyle\frac{1}{4\nu t}\left(1+\frac{At}{L_{0}^2}\right)}\right]
 \right\}
 \\
 -\displaystyle\frac 12 \left\{{\rm erf}\left[\displaystyle\frac{\sqrt{2}L_{0}
 -x}{\sqrt{4\nu t}}\right]+
 {\rm erf}\left[\displaystyle\frac{\sqrt{2}L_{0} +x}{\sqrt{4\nu t}}\right]
 \right\}
 \label{38}
 \end{array}
 \end{equation}
 The definition of the error function is:
 \begin{equation}
 {\rm erf}z =\frac{2}{\sqrt{\pi}}\int_{0}^{z}\exp(-t^2)dt.
 \label{39}
 \end{equation}
 The asymptotic behavior of the error function for large arguments is
 \begin{equation}
 {\rm erf}z =\pm \left(1-\frac{1}{\pi}e^{-z^2}\left[
 \frac{\Gamma(\frac{1}{2})}{z}-
 \frac{\Gamma(\frac{3}{2})}{z^3}+...\right]\right), \,
 z\rightarrow\pm\infty.
 \label{40}
 \end{equation}
 Using (\ref{40}) we find from (\ref{38}) for $|x|<\sqrt{2}L_{0}$, $t$ finite,
 $\nu/A\ll 1$:
 \begin{equation}
 U(x,t) \approx \displaystyle\frac{1}{\sqrt{1+\displaystyle\frac{At}{L_{0}^2}}}
 \exp\left(\frac{A}{2\nu}\left[1-\displaystyle\frac{x^2}{2L_{0}^2
 (1+\displaystyle\frac{At}{L_{0}^2})}\right]\right).
 \label{41}
 \end{equation}

 On the other hand, for
 $\sqrt{2}L_{0}<|x|<\sqrt{2}L_{0}\left(1+\displaystyle\frac{At}{L_{0}^2}
 \right)$ we have for $t$  finite and $\nu/A\ll 1$, $\nu t/L_0^2\ll 1$ and
 $At/L_{0}^2 \gg \nu/A$
 (the last inequality is necessary for the cancellation of the last two
 error functions in (36)):
 \begin{equation}
 U(x,t)\approx\frac{1}{\sqrt{1+\displaystyle\frac{At}{L_{0}^2}}}\exp
 \left(\frac{A}{2\nu}\left[1-\frac{x^2}{2L_{0}^2(1+\displaystyle\frac{At}
 {L_{0}^2})}\right]\right)+1.
 \label{42}
 \end{equation}

 For  $|x|<\sqrt{2A\left(t+\displaystyle\frac{L_{0}^2}{A}\right)}$, $t$ finite
 and $\nu/A\ll 1$ we can neglect the second term at the righthand side of
 (\ref{42}).
 Taking $\nu\rightarrow\ 0$ with $t$ finite in (\ref{41}) or (\ref{42})
 we obtain using (\ref{hc}):
 \begin{eqnarray}
 v(x,t) \approx \frac{x}{t}\displaystyle\frac{\displaystyle\frac{At}{L_{0}^2}}
 {1+\displaystyle\frac{At}
 {L_{0}^2}},\,|x|<x_s \nonumber\\
 v(x,t) =0,  \, |x|>x_s,
 \label{43}
 \end{eqnarray}
 where $x_s$ is the shock coordinate
 \begin{equation}
 x_{s}=\sqrt{2A\left(t+\frac{L_{0}^2}{A}\right)}.
 \label{44}
 \end{equation}

 However, for growing $t$-values the sharp discontinuity of the N-wave
 solution (\ref{43}) is broadened
 and we can calculate a shockwidth which depends on a small but finite value
 of $\nu$. Using (\ref{42}) in (\ref{hc}) and assuming that
 \begin{equation}
 t\gg \frac{L_{0}^2}{A},
 \label{45}
 \end{equation}
 we obtain
 \begin{equation}
 v(x,t)=\frac{x}{t}\frac{\displaystyle\frac{L_{0}}{\sqrt{At}}
 \exp\left[\frac{A}{2\nu}-\frac{x^2}{4\nu t}\right]}
 {\displaystyle\frac{L_{0}}{\sqrt{At}}\exp\left[\frac{A}{2\nu}-\frac{x^2}
 {4\nu t}\right]+1}.
 \label{46}
 \end{equation}
 Using (\ref{b}) and (\ref{34}) we find for the present case in (\ref{BRe})
 $L_{eff}=L_0\sqrt{4\pi\nu/A}$ and
 \begin{equation}
 B=\sqrt{\frac{4\pi L_{0}^2\nu}{A}}\exp\left[\frac{A}{2\nu}\right]=
 L_0\sqrt{\displaystyle\frac{2\pi}{Re_0}}\,e^{Re_0},
 \label{47}
 \end{equation}
 and thus, using (\ref{green}), we see that the result (\ref{46})
 is a special case of (\ref{apsol0}).

 From the expression (\ref{46}) we will now find the width of the shock.
 If we define the position $x_{s}$
 of the shock as the coordinate where the wave amplitude has decreased to
 half its maximal value, we
 find from (\ref{23}) the following expression for $x_{s}$:
 \begin{equation}
 x_{s} = \sqrt{2t\left(A-\nu\ln\frac{At}{L_0^2}\right)}.
 \label{48}
 \end{equation}
 Evaluating the expression (\ref{46}) for $x$ in the neighborhood of
 the shock, or more precisely, for
 \begin{equation}
 \frac{|x-x_{s}(t)|^2}{\nu t}\ll  1
 \label{49}
 \end{equation}
 we find
 \begin{equation}
 v(x,t)= \frac{x_{s}(t)}{2t}\left(1-\tanh\frac{x-x_{s}(t)}{\delta}\right),
 \label{50}
 \end{equation}
 where the shockwidth is given as
 \begin{equation}
 \delta = \frac{4\nu t}{x_{s}(t)}.
 \label{51}
 \end{equation}

 In order to decide which of the two waveforms (\ref{43}) and (\ref{50})
 is most appropriate we compare
 the expressions for $x_{s}$ (\ref{44}) and (\ref{48}).
 From these formulas we see directly that the shock
 velocity ($v_s=\displaystyle\frac{dx_{s}}{dt}$) decreases
 faster for growing $t$ with
 nonzero viscosity.
 The zero viscosity expression (\ref{44}) can be used as long as the
 correction to the N-wave shock position
 $x= \sqrt{2At}$ is much greater in (\ref{44}) than in (\ref{48}), which means
 \begin{equation}
 \frac{L_0^2}{At}\gg\frac{\nu}{A}\ln\left(\frac{At}{L_{0}^2}\right)
 \;\;, \;\;\;\;
 t\ln\frac{t}{t_{M,nl}}\ll \frac{A}{\nu}t_{M,nl},
 \label{52}
 \end{equation}
 where the nonlinear time $t_{M,nl}$ is
 defined as
 \begin{equation}
 t_{M,nl}=\frac{L_{0}^2}{A}.
 \label{53}
 \end{equation}
 In the notation $t_{M,nl}$, $M$ stands for the kind of pulse and $nl$
 stands for "nonlinear".

 The energy of a pulse is defined as the integral over the pulse length of
 the square of the fluid velocity.
 We will calculate the energies of some of the pulses studied above.
 For the N-wave (\ref{43}) we obtain
 \begin{equation}
 E(t) =\int v^2(x,t) dx =\frac{2^{5/2}}{3} \frac{A^2}
 {L_0 \left(1+\displaystyle\frac{At}{L_{0}^2}\right)^{1/2}}.
 \label{54}
 \end{equation}
 Thus for $t\gg t_{M,nl}$ and (\ref{52}) still valid we see from
 (\ref{54}) that $E(t)$ behaves like
 $A^{\frac{3}{2}}/t^{\frac{1}{2}}$ (\ref{E}) and thus depends on the
 initial scale only in the next order of $t$, i.e.  $\left(L_0^2
 A^{1/2}/t^{3/2}\right)$.

 The energy of the wave under the condition $t\gg t_{M,nl}$ but (\ref{52})
 no longer valid is obtained from (\ref{46}).
 We obtain
 \begin{equation}
 E(t)=\int\limits^{+\infty}_{-\infty}\left[\frac{x}{t}\frac
 {\displaystyle\frac{L_0}{\sqrt{At}}\exp\left(\frac{A}{2\nu}-\frac{x^2}
 {4\nu t}\right)}{1+\displaystyle\frac{L_0}{\sqrt{At}}\exp\left(
 \frac{A}{2\nu}-\frac{x^2}{4\nu t}\right)}\right]^2 dx.
 \label{55}
 \end{equation}
 After transformation of the integral in (\ref{55}) we obtain
 \begin{equation}
 E(t)=\frac{2\nu^{1/2}B^2}{\pi t^{3/2}}\int\limits^{\infty}_{0}
 \frac{\xi^{1/2}}{\left(e^{\xi}+\displaystyle\frac{B}{\sqrt{4\pi\nu t}}
 \right)^2} d\xi \;\;,
 \label{56}
 \end{equation}
 where $B$ is given in (\ref{47}).
 We need the formula (\cite{Prudnikov}, formula (2.3.13.27))
 \begin{equation}
 \int\limits^{\infty}_{0}\frac{x^{\alpha-1}e^{-px}}{\left(e^x-z\right)^2} dx=
 \Gamma(\alpha)\left[\Phi(z,\alpha,p+2)-(p+1)\Phi(z,\alpha,p+2)\right],
 \label{57new}
 \end{equation}
 where $\Phi$ is given by the power series
 \begin{equation}
 \Phi(z,\alpha,p)=\sum^{\infty}_{n=0}(p+n)^{-\alpha}z^{n}, \;\;|z|<1.
 \label{58new}
 \end{equation}
 Using (\ref{57new}), with $\alpha=3/2$ and $p=0$, in (\ref{56}) an energy
 expression, valid for $t\gg t_{M,nl}$ and (cf. (\ref{52}))
 $t\ln\displaystyle\frac{t}{t_{M,nl}}\gg\frac{A}{\nu}t_{M,nl}$, is obtained:
 \begin{equation}
 E(t)=\frac{2\nu^{1/2}B^2}{\pi t^{3/2}}\Gamma(3/2)\left[\Phi(-\frac{B}
 {\sqrt{4\pi\nu t}},\frac 12,2)-\Phi(-\frac{B}{\sqrt{4\pi\nu t}},\frac
 32,2)\right] \;\;.
 \label{59new}
 \end{equation}
 From (\ref{59new}) the energy of the wave in the linear region is obtained by
 keeping only the power zero in the series for $\Phi$:
 \begin{equation}
 E(t) = \frac{\sqrt{2}\nu^{3/2} L_{0}^2}{A}
 \displaystyle\frac{e^{A/\nu}}{t^{3/2}},
 \label{60new}
 \end{equation}
 which is a special case of (\ref{30}) with $B$ given by (\ref{47}). After
 introduction of the
 "linear" time $t_{M,lin}$, the validity condition of (\ref{60new}) is written:
 \begin{equation}
 \frac{B}{\sqrt{4\pi\nu t}}\ll 1,\;\;t\gg t_{Mlin}=t_{Mnl}\exp(A/\nu)=
 \exp(A/\nu)\frac{L_0^2}{A}=t_{Mnl}e^{2Re_0}\;\;.
 \label{61new}
 \end{equation}

 For the negative $A$ the pulse decays much faster than for $A>0$.
 For $\nu\to 0$ at $t>t_{Mnl}$ the initial pulse transforms into so called
 S-wave \cite{Whitham,GEP99}, and the energy becomes
 \begin{equation}
 E(t)=\frac{2^{5/2}}{3}\frac{L_0^3}{t^2}\;\;,
 \label{57}
 \end{equation}
 which is independent of the initial amplitude $|A|$ of the pulse. Of course
 the reason why the energy
 of the S-wave decreases faster with $t$ than the energy of the N-wave is
 that the length of the
 N-wave increases with growing $t$ in contrast to the S-wave.

 At large Reynolds number $|A|/2\nu\gg 1$ the constant $B$ (\ref{b})
 which determines the linear stage of evolution is
 \begin{equation}
 B=-2^{3/2}L_0
 \label{58}
 \end{equation}
 and much smaller than for positive $A$ with the same Reynolds number.
 The region of validity of linear regime is obtained from (\ref{bg2}) and
 (\ref{58}) as
 \begin{equation}
 t\gg L_0^2/\nu\;\;.
 \label{58a}
 \end{equation}
 Thus for the pulses with negative $A$ the linear stage begins much earlier
 than for the pulses with positive $A$, for which the linear time
 $t_{Mlin}$ (\ref{61new}) increases exponentionally  with the Reynolds number
 $Re_0=A/2\nu$.

 Nevertheless, we can see on this simple example the main
 nontrivial properties of the old-age behavior of the pulses in the
 case of large Reynolds number.

 In the limit of vanishing viscosity ($\nu\to 0$)
 the energy of the pulse (\ref{E}),(\ref{54}) and the shape of the
 pulse (\ref{wknsol}),(\ref{xs}),(\ref{43}) do not depend on the
 length of $L_0$ asymptotically for $t\rightarrow\infty$.
 They  are determined only by the
 maximum of initial potential $\psi_0$ (in our case -- of the area of
 triangular pulse of initial pulse $A$).
 Nevertheless, at the linear
 stage the energy depends not only of $A$, but is also proportional
 to $L_0^2$ (\ref{55}).
 We can remark that the energy of initial pulse is proportional to $A^2/L_0$.

 The other property that we stress is that the amplitude $B$
 (\ref{BRe}),(\ref{47}) of the wave at the old-age stage depends
 exponentially on the amplitude $A$ of the initial pulse.
 For small wave numbers $k$ the Fourier component $C(k,t)$ of the
 pulse depends linearly on $k$: $C(k,t)\sim b(t)k$.
 At the linear stage the slope $b(t)$ of the Fourier component does
 not depend on time: $b(t)\simeq B=const$.
 At the nonlinear stage the growing of the slope until the very large
 linear time (\ref{61new}) leads to an exponentionally large value (in
 $1/\nu$) of $b(t)$ at the linear stage (cf.(\ref{47})).

 \subsection{The decay of an N-wave with random amplitude}

 Let us discuss the evolution of N-wave (\ref{35}) with random initial
 amplitude.
 We assume that the amplitude $A$ in (\ref{33}) has a gaussian probability
 distribution function with a zero mean value
 \begin{equation}
 W(A)=\frac{1}{\sqrt{2\pi\sigma^2_{\psi}}}e^{-A^2/2\sigma^2_{\psi}} \;\;.
 \label{59}
 \end{equation}
 From (\ref{35}) it is easy to see that mean initial field
 $\left<v_0(x)\right>$ is zero, and relative fluctuation of the
 energy (\ref{54}) at $t=0$ is
 \begin{equation}
 \delta E(0)=\frac{\left<E^2(0)\right>-\left<E(0)\right>^2}
 {\left<E(0)\right>^2}=2 \;\;.
 \label{60}
 \end{equation}

 In the previous section it was shown that there is strong difference
 between the
 decay of the pulse with positive and negative amplitude $A$.
 If we introduce $\bar E_+(t)$ -- the mean energy of pulse with positive
 amplitude $A$ (N-wave), and $\bar E_-(t)$ -- the mean energy of pulse
 with negative amplitude $A$ (S-wave), then at time
 \begin{equation}
 t\gg t_{M,nl}\simeq L_0^2/\sigma_{\psi}
 \label{61}
 \end{equation}
 from (\ref{54}), (\ref{57}) and (\ref{59}) we have
 \begin{equation}
 \bar E_{+}(t)/\bar E_{-}(t)\simeq (\sigma_{\psi}t/L_0^2)^{3/2}
 \simeq(t/t_{M,nl})^{3/2}\gg 1.
 \label{62}
 \end{equation}
 This fast decrease of pulses with negative $A$ will lead to the generation
 of a field with
 positive mean value $\langle v(x,t)\rangle$ from an initial field $v_0(x)$
 with zero mean value
 $\langle v_0(x)\rangle=0$.

 At time $t\gg t_{M,nl}$ (\ref{61}) we can neglect the influence of the
 negative pulse on the mean field and use the expressions
 (\ref{wknsol}),(\ref{xs}) for the velocity.  In different
 realizations we will have the N-wave with the same slope $v'_x=1/t$
 and random position of the shocks.

 Let us introduce the cumulative probability of the random amplitude
 \begin{equation}
 Q_H(H)=Prob(A<H)=\int\limits_{-\infty}^{H}W(A)dA \;\;.
 \label{63}
 \end{equation}
 Then from (\ref{wknsol},\ref{xs})
 \begin{equation}
 \left\{
 \begin{array}{l}
 v(x,t)=x/t \;\; {\rm with \;probability}\;\; 1-Q_H(x^2/2t)\\
 v(x,t)=0 \;\;\;\;\; {\rm with \;probability} \;\; Q_H(x^2/2t)\\
 \end{array}
 \right.
 \label{64}
 \end{equation}
 From (\ref{64}) we obtain for the mean velocity $\langle v(x,t)\rangle$
 and its variance $\sigma^2_v(x,t)=\langle(v(x,t)-\langle v(x,t)\rangle)^2
 \rangle$ the following expressions:
 \begin{equation}
 \langle v(x,t)\rangle=\displaystyle\frac xt \left( 1-Q_H\left(\frac{x^2}{2t}
 \right)\right)\;\;,
 \label{65}
 \end{equation}
 \begin{equation}
 \sigma^2_v(x,t)=\displaystyle\frac{x^2}{t^2}Q_H\left(\frac{x^2}{2t}\right)
 \left(1-Q_H\left(\frac{x^2}{2t}\right)\right) .
 \label{66}
 \end{equation}
 For the Gaussian distribution of $A$ (\ref{59}) we have from (\ref{63}),
 (\ref{65})
 \begin{equation}
 \langle v(x,t)\rangle =\displaystyle\frac{x}{2t}\left(1-{\rm erf}\left(
 \frac{x^2}{2\sqrt 2 \sigma_{\psi}t}
 \right)\right) ,
 \label{67}
 \end{equation}
 \begin{equation}
 \sigma^2_v(x,t)=\displaystyle\frac{x^2}{4t^2}
 \left(1-{\rm erf}^2\left(\frac{x^2}{2\sqrt 2 \sigma_{\psi}t}\right)\right),
 \label{68}
 \end{equation}
 where ${\rm erf}(z)$ is the error function (\ref{39}).

 At small $x<(\sigma_{\psi}t)^{1/2}$ the mean field $\langle v(x,t)
 \rangle$ is half the regular field with
 $A\sim\sigma_{\psi}$, due to the fact that only in that half of
 realizations, in which $A>0$, we have a relatively slowly decaying
 N-wave.
 The mean field $\langle v(x,t)\rangle$ (\ref{67}) has not a clear
 stressed shock front and has a very fast decaying tail at
 $x\gg\sqrt{\sigma_{\psi}t}$:
 \begin{equation}
 \langle v(x,t)\rangle\simeq\displaystyle\frac{\sqrt 2}{\sqrt\pi}\frac{\sigma_
 {\psi}}{x}\exp\left[-\frac{x^4}{8\sigma^2_{\psi}t^2}\right].
 \label{69}
 \end{equation}
 It is easy to see that both mean field $\langle v(x,t)\rangle$ and variance
 $\sigma^2_v(x,t)$ have self-similar behaviour
 \begin{equation}
 \langle v(x,t)\rangle =\displaystyle\sqrt{\frac{\sigma_{\psi}}{t}}
 \bar V\left(\frac{x}{\sqrt{\sigma_{\psi}t}}\right),\;\;\;\;
 \sigma^2_v(x,t)=\displaystyle\frac{\sigma_{\psi}}{t}\bar{\sigma}^2_V
 \left(\frac{x}{\sqrt{\sigma_{\psi}t}}\right),
 \label{70}
 \end{equation}
 where $\bar V$, $\bar{\sigma}_V$ are given as:
 \begin{equation}
 \bar V(y)=\frac 12 y\left(1-{\rm erf}\left(\frac{y^2}{2\sqrt 2}\right)\right),
 \;\;\;\;
 \bar{\sigma}^2_V(y) = \frac 14 y^2\left(1-{\rm erf}^2\left(\frac{y^2}
 {2\sqrt 2}\right)\right).
 \label{71}
 \end{equation}
 Due to the self-similarity the relative integral fluctuation of the field
 \begin{equation}
 \delta v^2 =\displaystyle\frac{\int\limits^{+\infty}_{-\infty}
 \sigma_v^2(x,t)dx}{\int\limits^{+\infty}_{-\infty}\langle v(x,t)\rangle^2 dx}=
 3.356
 \label{72}
 \end{equation}
 does not depend on $t$ and is of the order of unity.

 The situation radically changes on the linear stage when the field is
 described by the equation (\ref{vapprox}).
 At this stage we have the self-similar evolution of the field
 \begin{equation}
 v(x,t)=\frac{B}{t}\bar V_{lin}\left( \frac{x}{\sqrt{2\nu t}}\right),
 \label{73}
 \end{equation}
 where $\bar V_{lin}(y)$ is defined as
 \begin{equation}
 \bar
 V_{lin}(y)=\frac{y}{\sqrt{2\pi}}\exp\left[-\frac{y^2}{2}\right],
 \label{73a}
 \end{equation}
 and $B$ is a random amplitude (\ref{47}) with nonzero mean value.
 For the gaussian distribution of $F$ in (\ref{31}) we have from (\ref{b})
 \begin{equation}
 \langle B\rangle=\int\limits^{+\infty}_{-\infty}\left(
 e^{M^2(x)Re_0^2/2}-1\right)dx.
 \label{74}
 \end{equation}
 Here we introduce the effective Reynolds number $Re_0$
 \begin{equation}
 Re_0=\langle F^2(x)\rangle^{1/2}/2\nu=\sigma_{\psi}/2\nu.
 \label{75}
 \end{equation}
 From (\ref{74}) we see that mean value $\langle B\rangle$ does not depend
 on the fine structure of the carrier $f(x)$ and is always positive.
 The positive mean value $\langle B\rangle$ is a result of generation of
 mean field at the nonlinear stage.
 Here we consider the case of large
 Reynolds number, where we can use the asymptotic expression for $B$
 (\ref{47}).
 The n-th moment of $B$ will be expressed through the  probability
 distribution function (p.d.f.) of $A$ (\ref{59}):
 \begin{equation}
 \langle B^n\rangle= (4\pi
 L_0^2\nu)^{n/2}\int\frac{e^{nA/2\nu}}{A^{n/2}} W(A)dA.
 \label{76}
 \end{equation}
 Using the steepest descent method for $Re_0\gg 1$ we have from
 (\ref{59}), (\ref{76})
 \begin{equation}
 \langle B^n\rangle=\left(\frac{2\pi}{n}\right)^{n/2}L_0^n\frac{1}{Re_0^n}
 e^{n^2 Re_0^2/2}.
 \label{77}
 \end{equation}
 Thus we have very fast growing of the momentum with increasing  $n$.
 From (\ref{73}) one can see that at the linear stage both mean field
 and variance are self-similar, and that the relative integral
 fluctuation $\delta v^2$ does not depend on $t$.
 But in contrast to the
 nonlinear stage (\ref{72}), in the linear stage the relative
 integral fluctuation of the field $\delta v^2$ is
 extremely large for large Reynolds number:
 \begin{equation}
 \delta v^2=\frac{\langle B^2\rangle - \langle B\rangle^2}
 {\langle B\rangle^2}\simeq \frac 12 e^{Re_0^2}.
 \label{78}
 \end{equation}

 We have  similar behaviour also for the relative
 fluctuation of the energy $\delta E(t)$ (\ref{60}).
 At the
 nonlinear stage at $\sigma_{\psi}t\gg L^2_0$ we have from
 (\ref{E}), (\ref{59}), (\ref{60})
 \begin{equation}
 \delta E(t)=\frac{2\sqrt{\pi}}{\Gamma^2(5/4)}-1\simeq 3.376 \;\;.
 \label{79}
 \end{equation}

 Thus at the nonlinear stage $\delta E(t)$ does not depend on the
 initial scale $L_0$  and the variance $\sigma_{\psi}$ of the p.d.f.
 of amplitude $A$.

 At the linear stage the fluctuation of the energy is determined by
 the fourth and the second moment of $B$, and from (\ref{77}) we have
 also a strong enhancement of energy fluctuation (\ref{79})
 \begin{equation}
 \delta E(t)=\frac{1}{4}e^{4Re_0^2}.
 \label{79a}
 \end{equation}

 \section{The evolution of a pulse with monochromatic carrier}

 In this section we consider the evolution of a pulse with monochromatic
 carrier
 \begin{equation}
 \psi_0(x)=M(x)A\,{\rm cos}\,k_0x, \;\;\;\;
 v_0(x)\simeq M(x)a_0\, {\rm sin}\, k_0x.
 \label{80}
 \end{equation}
 Here $A=a_0/k_0$, and we
 assume that the length scale $L_0$ of the modulation function $M(x)$ is
 much greater then the period $l_0$ of the carrier ($l_0=2\pi/k_0$).
 Below we consider two
 large parameters $Re_0=A/2\nu$ and $L_0/l_0$.
 For different ratios
 of these parameters we have different regimes of wave evolution.

 The pure monochromatic signal $(M(x)\equiv 1)$ is characterized by
 the nonlinear time $t_{s,nl}=1/a_0k_0$ ("$s$" stands for "signal") and the
 linear time
 $t_{s,lin}=1/\nu k_0^2$ \cite{GMS91}.
 At $t_{s,nl}\ll t\ll t_{s,lin}$ ($Re_0\gg 1$) the monochromatic wave transforms
 into the sawtooth wave with the slope $v'_x=1/t$.
 The shock amplitude $\triangle V =l_0/2t$, as well as the energy of the wave
 $E=l_0^2/12 t^2$, does not depend on the initial amplitude.
 The shockwidth
 \begin{equation}
 \delta =2\nu/\triangle V=4\nu t/l_0
 \label{81}
 \end{equation}
 increases with time, and is, at $t\sim t_{s,lin}$,  of the same order as a
 period.
 For $t\gg t_{s,lin}$ we have the linear regime of evolution, where the
 wave form is sinusoidal again with exponentionally decaying amplitude
 $v(x,t)=2\nu k_0 \exp[-\nu k_0 t^2]{\rm sin}\, k_0 x$.

 For a pulse with monochromatic carrier a large-scale component
 $v_l(x,t)$ is generated.
 The interaction of the high frequency
 component with the large-scale wave $v_l(x,t)$ will change the
 evolution of the carrier.

 \subsection{The nonlinear stage of evolution at large Reynolds number}

 Below we give a short summary of pulse evolution at $\nu\to 0$ based on
 the paper \cite{GEP99}, and discuss the influence of finite dissipation
 on the evolution of large scale and high frequency carrier.

 The nonlinear effect leads to the generation of the large-scale
 component $v_l(x,t)$, and at $t\gg t_{s,nl}$ this component obtains
 the stationary form
 \begin{equation}
 v_l(x,t)=-M'(x)\,A,
 \label{82}
 \end{equation}
 which is equal to the form of simple pulse $v_0(x)=M'(x)A$ with the
 same initial potential $\psi_0(x)=M(x)\,A$.
 The evolution of the large-scale component is characterized by the
 nonlinear time $t_{M,nl}=L_0^2/A$ (\ref{53}), and (\ref{82}) is
 valid for $t\le t_{M,nl}$.
 At time
 \begin{equation}
 t_{int}=\frac{L_0 l_0}{A}
 \label{83}
 \end{equation}
 the energy of large-scale and
 small-scale components are of the same order.
 At $t\gg t_{s,nl}$ the
 evolution of the large-scale component is equal to the evolution of
 a simple pulse with $\psi_0(x)=M(x)\,A$.

 For $M(x)=(1-x^2/2L_0^2),\;|x|<\sqrt 2 L_0$ the evolution
 of the large-scale component is described by the expressions
 (\ref{43}), (\ref{44}), (\ref{54}).
 At $t\gg t_{s,nl}$, in the limit
 $\nu\to 0$, the evolution of the large-scale component is determined
 by only one parameter $A=a_0/k_0$ of the initial perturbation.

 At $t\gg t_{s,nl}$ the amplitudes $\triangle V$ of the shocks of the
 small-scale
 components do not depend on the initial amplitude: $\triangle V=l_0/2t$,
 but due to the interaction with large-scale component they have nonzero
 velocity $V_r\simeq-A\,M'(x_r)=v_l(x_r)$ (\ref{82}), where $x_r$ is the
 initial position of the shock.
 The distance between the shocks increases with time as
 \begin{equation}
 l(t)=l_0(1+At/L_0^2).
 \label{84}
 \end{equation}

 The collision of the shocks of small-scale components moving with
 constant velocity of fine structure with shocks of the large-scale
 structure (\ref{44}) leads to  decrease of the number of
 shocks (see fig. 3 from \cite{GEP99}).
 The last collision occurs at time
 $t=\displaystyle\frac{8L_0^2}{l_0^2}t_{M,nl}$, and after this time a
 pure N-wave remains.

 At finite Reynolds number the width of the shocks of the large-scale
 component increases with time (\ref{81}).
 The linear spreading of small structure leads to the increase of
 the distance between the shocks (\ref{84}).
 Comparing (\ref{81}) with (\ref{84}) we find that, if the initial
 Reynolds number satisfies the condition
 \begin{equation}
 \frac{A}{2\nu}=Re_0\gg \frac{L_0^2}{l_0^2},
 \label{85}
 \end{equation}
 then the nonlinear structure of shocks will be conserved.
 This is because the relative shock width is
 \begin{equation}
 \frac{\delta (t)}{l(t)}=4\frac{\nu}{A}\,\frac{L_0^2}{l_0^2}.
 \label{86}
 \end{equation}
 In the opposite case, under the condition
 \begin{equation}
 1\ll Re_0 \ll L_0^2/l_0^2
 \label{87}
 \end{equation}
 the nonlinear structure will dissipate before the nonlinear
 distortion of large-scale component begins.

 The evolution of the large-scale component at finite Reynolds
 number will be described by the same equations as the
 evolution of a simple wave (\ref{49}) - (\ref{51}).
 Only the position of the shocks $x_s(t)$ will be determined
 by some other equation (\ref{23}), where $L_{eff}$ depends
 on the fine structure of the initial pulse.
 This dependence leads to the sensitivity of the old-age behaviour on
 the fine structure of the initial pulse.

 \subsection{Old-age linear stage evolution of pulse with
 monochromatic carrier}

 The final linear stage of evolution is described by the equation
 (\ref{vapprox}), where the constant $B$ (\ref{b}) is now given by
 \begin{equation}
 B=\int\limits^{+\infty}_{-\infty}\{\exp\left[ Re_0 M(x){\rm cos}\, k_0
 x-1\right]dx\}.
 \label{88}
 \end{equation}
 At large Reynolds number $Re_0=A/2\nu$ the constant $B$ may be written in
 the form $B=L_{eff}\exp[Re_0]$ (\ref{BRe}).
 In fact for the large Reynolds number the main contribution in the integral
 in (\ref{88}) comes from the neighborhood of points $y_r=l_0 r \;(l_0=
 2\pi/k_0, \; r=0,\pm 1,\pm 2,...)$.
 An evaluation of $B$ by the steepest descent method then gives
 \begin{equation}
 B=\sum_r \frac{\sqrt{2\pi}}{k_0\sqrt{Re_0 M(y_r)}}\exp[Re_0 M(y_r)].
 \label{89}
 \end{equation}
 From (\ref{89}) we see that the prefactor $L_{eff}$ in (\ref{BRe}) strongly
 depends on the ratio of two large numbers: the Reynolds number $Re_0$
 and the number $L_0^2/l_0^2$ $(l_0=2\pi /k_0)$.
 When the relation (\ref{85}) is valid, then only the first term $r=0$ in
 (\ref{89}) is significant, and from (\ref{BRe}), (\ref{89}) we have
 \begin{equation}
 L_{eff}=\frac{l_0}{\sqrt{2\pi Re_0}}.
 \label{90}
 \end{equation}

 In the case of moderate Reynolds number, when the condition (93) is valid,
 we have from (\ref{BRe},\ref{89})
 \begin{equation}
 L_{eff}=\frac{L_0}{Re_0}.
 \label{91}
 \end{equation}

 The results above should be compared with expression (\ref{47}) of $B$
 for the simple pulse for which we have $L_{eff}=L_0\sqrt{2\pi /Re_0}$.
 One can see that the modulation leads to  faster transformation of
 the wave into the linear regime of evolution and consequently to
 decrease of the amplitude of the wave ($B\sim L_{eff}\exp[A/2\nu]$)
 at the old-age stage.

 \section{The evolution of a pulse with noise carrier}

 In this section we will study the evolution of  a pulse with noise carrier
 $f(x)=-F'(x)$ (\ref{32}).
 We assume that the potential $F(x)$ is homogeneous gaussian noise with
 rapidly decreasing covariance function
 \begin{equation}
 B_{\psi}(\rho)=\langle F(x)F(x+\rho)\rangle=\sigma^2_{\psi}R(\rho)=
 \sigma^2_{\psi}\left( 1-\frac{\rho^2}{2l_0^2}+\frac{\rho^4}{4l_1^2}+...
 \right).
 \label{92}
 \end{equation}

 In the limit of vanishing viscosity the continuous homogeneous field
 $v_0(x)=f(x)$ transforms into sequence of sawtooth pulses with equal slope
 $v'_x=1/t$ and with random position of shocks.
 Due to the collision and merging of the shocks, their number decreases and the
 characteristic scale of random field increases.
 This effect makes all the statistical properties of the field self-similar
 and they are determined by only one scale $l(t)$, which can be interpreted as
 characteristic distance between the zeroes of $v(x,t)$ or between the shocks
 \cite{GMS91,MSW95,GSAFT97}.

 The evolution of the integral scale $l(t)$ in time due to merging of the
 shocks (\cite{GMS91}, see eq. 4.15, p. 170)
 \begin{equation}
 l(t)=(\sigma_{\psi}t)^{1/2}{\rm ln}^{-1/4}\left(\frac{t\sigma_v^2}
 {2\pi\sigma_{\psi}}\right)
 \label{93}
 \end{equation}
 depends on only two integral characteristics of the initial
 homogeneous field
 \begin{equation}
 \sigma^2_{\psi}=\langle F^2(x)\rangle, \;\;
 \sigma^2_v=\sigma^2_{\psi}/l_0^2= \langle f^2(x)\rangle.
 \label{94}
 \end{equation}
 Here $l_0$ is the correlation length of the initial potential.
 Due to the merging of the shocks the energy density of the random
 wave $\langle v^2(x,t)\rangle=l^2(t)/t^2\sim t^{-1}$ decreases
 slower then the energy of harmonic perturbation.

 In case of the finite Reynolds number the thickness of the shock in
 the sawtooth wave originating from a monochromatic wave is given in
 (\ref{81}).
 We have the same expression for a random wave as well, where
 $\triangle V$ is the random amplitude of the shock.
 For the estimations we can assume $\triangle V\sim l(t)/t$.

 The ratio between the integral scale $l(t)$ and internal scale
 $\delta(t)$ is the effective Reynolds number $Re(t)$:
 \begin{equation}
 Re(t)\approx\frac{l(t)}{\delta(t)}\approx\frac{l(t)\triangle V(t)}{\nu}
 \sim Re_0 \ln^{-1/2}\left(\frac{t}{t_{nnl}}\right),
 \label{95}
 \end{equation}
 where $Re_0$ and $t_{n,nl}$ ("$n$" stands for "noise") are defined as
 \begin{equation}
 Re_0=\sigma_{\psi}/2\nu , \;\; t_{n,nl}=l_0^2/\sigma_{\psi}.
 \label{96a}
 \end{equation}

 Thus the effective Reynolds number $Re(t)$ logarithmically slowly
 decreases with time, and the linear stage of evolution starts at very large
 time $t_{n,lin}\simeq t_{n,nl}\exp(Re_0^2)\gg t_{n,nl}$, $(Re(t_{n,lin})\sim
 1)$.
 At this time the nonlinear effects become unimportant and the noise enters
 into the linear mode where its damping is mainly determined by linear
 dissipation.
 On the base of the solution (\ref{uhc}), as it was shown in
 \cite{GMS91}, the energy decays as
 \begin{equation}
 \sigma^2_v(t)=(\nu l_0^2/t^3)^{1/2} Re_0^{1/2} e^{Re_0^2}.
 \label{96}
 \end{equation}

 At the linear stage the distribution of the random field converges
 to the distribution of the homogeneous Gaussian field with zero mean
 velocity and variance according to (\ref{96}) \cite{AMS94}.

 The evolution of the pulse with noise carrier in the limit of
 vanishing viscosity $(\nu\to 0)$ was considered in the paper
 \cite{GEP99}.
 It was shown that an initial perturbation $v_0(x)$
 transforms to an N-wave.
 In the case, when the scale of the carrier
 $l_0$ is much smaller then scale of modulation function $L_0$,
 it was shown that the
 fluctuation of the shock positions is relatively small.

 Below we consider the properties of the pulse with noise carrier at
 the nonlinear stage for finite Reynolds number and the old-age behaviour
 of the pulse.

 \subsection{Nonlinear stage of evolution of pulse with noise carrier}

 In the case of vanishing viscosity we can introduce two characteristic times:
 the nonlinear time for noise carrier evolution $t_{n,nl}=l_0^2/\sigma_{\psi}$
 and nonlinear time for modulation evolution $t_{M,nl}=L_0^2/\sigma_{\psi}$.
 At $t\gg t_{n,nl}$ the initial wave transforms into the sequence of sawtooth
 pulses, with the integral scale $l_M(x,t)\simeq M^{1/2}(x)
 l_0(t/t_{n,nl})^ {1/2}$ and the energy density $\bar E
 (x,t)=l_M^2(x,t)/t^2\simeq M(x) \sigma^2_v(t/t_{n,nl})^{-1}$
 depending slowly on the coordinate \cite{GEP99}.
 At this stage the
 nonlinearity leads to partial depression of modulation and
 to generation of the mean field $v_l(x,t)=\langle v(x,t)\rangle=
 -M'(x)\sigma_{\psi}
 \left( \ln M(x)t/2\pi t_{nnl}\right)^{1/2}$ \cite{GEP99}.
 Due to the merging of the shocks their number decrease, and at $t\gg t_{M,nl}$
 the initial pulse with noise carrier transforms into an N-wave with
 random positions of zero and shocks.
 The position of the N-wave zero is
 localized in an narrow region $\triangle l \simeq L_0/(\ln \,
 L_0^2/2\pi l_0^2)^{1/2}$ near the center of the initial pulse
 \cite{GEP99}.
 The position of the shocks are determined by the
 equation (\ref{xs}), where $A=\psi_m$ is the value of absolute
 maximum of the initial potential $\psi_0(x)$ (\ref{31}).
 It was shown that the cumulated probability
 $Q_H(H)={\rm Prob}(\psi_m<H)$ has the form \cite{GEP99}
 \begin{eqnarray}
 Q_H(H)=\exp[-N_{\infty}(H)],\nonumber\\
 N_{\infty}(H)=\left(\frac{\sigma^2_{\psi}}{H^2}\,\frac{L_0^2}{2\pi l_0^2}
 \right)^{1/2}\,\exp\left[-\frac{H^2}{2\sigma^2_{\psi}}\right].
 \label{97}
 \end{eqnarray}
 Here $N_{\infty}(H)$ is the mean number of intersections of level $H$
 by the initial potential $\psi_0(x)$ in the interval $|x|<L_*$ (\ref{inpot}),
 $L_0$ is a characteristic scale of the modulation function $M(x)$
 (\ref{34}) $(L_0\sim L_*)$,
 $l_0=\sigma_{\psi}/\sigma_v$ is an integral scale of the Gaussian
 homogeneous carrier potential $F(x)$ and $\sigma_{\psi}$ is its
 variance (\ref{94}).

 The mean velocity $\langle v(x,t)\rangle$, with $v(x,t)$ given by
 (\ref{64}) and its variance $\sigma^2_v(x,t)$ are described by
 equations (\ref{65}), (\ref{66}), where now the cumulative probability
 $Q_H(H)$ is determined by equation (\ref{97}), and not by the error
 function in (\ref{67}).
 It is easy to see from (\ref{70})
 that both mean field and variance have self-similar behaviour.
 From (\ref{65}) , (\ref{66}), (\ref{97}) we have (cf. (74))
 \begin{equation}
 \bar V (y,N)=y(1-q_N(y)), \;\; \bar{\sigma}^2_V(y,N)=y^2
 q_N(y)(1-q_N(y)),
 \label{98}
 \end{equation}
 \begin{equation}
 q_N(y)=Q_H(y^2\sigma_{\psi}/2)=\exp\left[-\frac{\sqrt 2
 N}{y^2}\,\exp\left(-\frac{y^4}{4} \right)\right],
 \label{99}
 \end{equation}
 where $y=x/\sqrt{\sigma_{\psi}t}$, and $N$ is a large parameter,
 proportional to the number of correlation lengths on the whole
 extension of the pulse:
 \begin{equation}
 N=\frac{L_0}{\sqrt {2 \pi}l_0}.
 \label{100}
 \end{equation}

 Contrary to the case of the simple initial pulse (\ref{71}), in the
 case of pulse with noise carrier the mean field and its variance
 have two different scales for $N\gg 1$ (see
 figs. 1 and 2). The two scales in the "b" and "c" curves in the figures are
 the width of the shock
 and the length of the pulse. In the "a" curves only the length of the pulse
 remains.
 For $N\gg 1$ the function
 $q_N(y)\simeq 0$ at $y<y_*\simeq (4\ln N)^{1/4}$.
 The function $q_N(y)$ increases rather fast to $1$ in the narrow
 region $(y-y_*)/y_*\simeq (4\ln N)^{-1}$.

 Thus for $N\gg 1$ the mean field has the N-wave similar structure with the
 dimensionless shock position
 \begin{equation}
 y_s=y_*=(4\ln N)^{1/4}=\left(2\ln\frac{L_0^2}{2\pi l_0^2}\right)^{1/4}.
 \label{101}
 \end{equation}
 The relative width
 of the shock of the mean field is
 \begin{equation}
 \frac{\triangle x_s}{x_s}=\frac{\triangle y}{y_*}\simeq\frac{1}{4\ln N}=
 \frac{1}{2\ln(L_0^2/2\pi l_0^2)}.
 \label{102}
 \end{equation}
 In the neighborhood of the shock position we can introduce a new variable z:
 \begin{equation}
 y=y_*(1+z/y_*^4),
 \label{103}
 \end{equation}
 and from (\ref{99}) we see that the shape of the front is described
 by double exponential distribution
 \begin{equation}
 q_N(y)=q_d((y-y_*)y_*^4), \;\; q_d(z)=e^{-e^{-z}}.
 \label{104}
 \end{equation}

 The variance $\sigma^2_v$ is different from zero in a narrow region
 (\ref{102}) near the shock position $y_*$ and for the relative
 integral fluctuation of the field (\ref{72}) we have from
 (\ref{98}), (\ref{99})
 \begin{equation}
 \delta v^2\simeq\frac{\triangle y}{y_*}\simeq\frac{1}{4\ln N}\ll 1.
 \label{105}
 \end{equation}
 A finite Reynolds number leads to a finite width of the shock.
 The shock structure in each realization is described by the
 expressions (\ref{50}), (\ref{51}).
 The shifting of the shock position
 $x_s$ from the zero viscosity limit $x_s=\sqrt{2\psi_m t}$
 $(A\equiv \psi_m)$ is given in (\ref{23}).
 At the nonlinear stage we can neglect this shifting.
 Then in each
 realization we have self-similar evolution of the pulse, and the
 relative width of the shock $\delta/x_s=2\nu/\psi_m$ does not depend
 on time.
 While for $N\gg 1$ ($L_0\gg l_0$) the maximum $\psi_m$ of
 the initial potential is localized in the narrow region $\triangle
 H/H_0\simeq(\ln N)$ near $H_0\simeq \sigma_{\psi}(2\ln N)^{1/2}$
 (\cite{GEP99} formulas (107), (113)), we can estimate the relative
 width of shock front as
 \begin{equation}
 \frac{\delta}{x_s}\simeq\frac{2\nu}{H_0}\simeq\frac{1}{Re_0}\,\frac{1}
 {(\ln N)^{1/2}},\;\; Re_0=\frac{\sigma_{\psi}}{2\nu}.
 \label{106}
 \end{equation}
 Comparing (\ref{106}) with (\ref{102}) we see that the influence of finite
 viscosity on the mean field is unimportant if
 \begin{equation}
 \left(\ln\frac{L_0^2}{2\pi l_0^2}\right)^{1/2}\simeq
 (\ln N)^{1/2}\ll Re_0.
 \label{107}
 \end{equation}
 In the opposite case,
 for extremely large ratio $L_0/l_0$, the width of the mean field
 will be determined by the viscosity.

 The displacements of shock position $x_s$ (\ref{23}) from the vanishing
 viscosity position $x_s=\sqrt{2\psi_m t}$ leads finally to the
 depressing of nonlinear effects.
 For $t\gg t_{lin}\sim L_{eff}^2 e^{\psi_m/\nu}/\nu$ (cf.(\ref{recond}))
 we have the linear stage of evolution.
 While for the random
 perturbation we have a large fluctuation of $t_{lin}$, the
 cumulative action of nonlinearity, which is proportional
 to $t_{lin}$, leads to strong fluctuation of the field at the
 linear stage.

 \subsection{Old-age linear stage of evolution of a pulse with noise carrier}

 At the old-age stage, when  the evolution of the pulse is described by
 linear solution (\ref{vapprox}) all the properties of the wave are
 determined by the constant $B$, given in (\ref{b}).
 The potential $\psi_0(x)$ (\ref{31}) is a random Gaussian function.
 The mean value of $B$ is given by the equation (\ref{74}), and
 does not depend on the fine structure of the carrier.
 For $Re_0=\sigma_{\psi}/2\nu\gg 1$ we have from (\ref{77})
 \begin{equation}
 \langle B\rangle=L_0\sqrt{2\pi}\frac{1}{Re_0}\, e^{Re_0^2/2}.
 \label{107a}
 \end{equation}

 Let us first consider the case of relatively large Reynolds number, when the
 integral (\ref{b}) may be calculated using the steepest descent
 method, and only the contribution of the absolute maximum is
 important:
 \begin{equation}
 B=\sqrt{\frac{4\pi\nu}{|\psi''_{m}|}}\, e^{\psi_m/2\nu}.
 \label{108}
 \end{equation}
 Here $\psi_m=\psi_0(x_m)$ is the value of absolute maximum of $\psi_0$,
 and $\psi''_m=\psi''_0(x_m)$ is the second derivative of
 the potential at this point.
 To find the statistical properties of
 $B$ we need to know the joint probability distribution of $\psi_m$
 and $\psi''_m$.
 From (\ref{92}) it is easy to find the correlation
 coefficient for $\psi_m$ and $\psi''_m$: $r=-l_1^2/l_0^2$.
 If we consider the conditional probability distribution function
 $W(\psi''_m/\psi_m)$, we then obtain for the conditional mean
 value $\langle \psi''_m\rangle_{\psi_m}$ and variance
 $\left(\sigma^2_{\psi''_m}\right)_{\psi_m}$:
 \begin{eqnarray}
 \langle \psi''_m\rangle_{\psi_m}=-\psi_m r\sigma^2_{\psi ''_m}
 /\sigma^2_{\psi_m}=-\psi_m/l_0^2,\nonumber\\
 \left(\sigma^2_{\psi''_m}\right)_{\psi_m}=\sigma^2_{\psi_m}(1-l^4_1/l^4_0)/
 l^4_1.
 \label{109}
 \end{eqnarray}
 While $B$ increases exponentionally with $\psi_m$ only
 the maximum $\psi_m \gg\sigma_{\psi}$ gives a significant
 contribution to the average characteristics of $B$.
 Thus in (\ref{108}) we can substitute $\langle\psi ''_m\rangle_{\psi_m}$
 (\ref{109}) instead of $\psi ''_m$, and so we have
 \begin{equation}
 B=l_0\sqrt{\frac{4\pi\nu}{\psi_m}}\, e^{\psi_m/2\nu}.
 \label{110}
 \end{equation}

 From (\ref{97}) we have for the probability distribution function of the
 absolute
 maximum of the initial potential $\psi_0(x)$:
 \begin{equation}
 W_H(H)=\frac{\partial Q_H}{\partial H}=-N'_{\infty}(H)e^{-N_{\infty}(H)}.
 \label{111}
 \end{equation}
 This function is localized near
 \begin{equation}
 H_0=\sigma_{\psi}(2\ln N)^{1/2}
 \label{112}
 \end{equation}
 and has a tail for $H\rightarrow\infty$:
 \begin{equation}
 W_H(H)=\frac{N}{\sigma_{\psi}}e^{-H^2/2\sigma^2_{\psi}}.
 \label{114}
 \end{equation}

 For the calculation of the moments of $B$ we can use the steepest
 descent method using the tail of the probability distribution function
 (\ref{114}):
 \begin{equation}
 \langle B^n\rangle=L_0\,l_0^{n-1}\left(\frac{2\pi}{n}\right)^{n/2}
 \frac{1}{Re_0^n}e^{n^2Re^2_0/2}.
 \label{115}
 \end{equation}
 We have the main contribution for the n-th moment (\ref{115})
 at the point $H_{*,n}\simeq n\sigma_{\psi}/2\nu=n\,Re_0$.
 Comparing $H_{*,n}$ with $H_0$ given in (\ref{112}) we see that the
 inequality (\ref{107}) is necessary for (\ref{115}) to be valid.

 Comparing (\ref{115}) with the relation (\ref{77}) for the moment of $B$
 for a simple pulse we see that for the small-scale noise carrier
 the moments $n\ge 2$ depend on the integral scale $l_0$ of the
 carrier.

 The relative integral fluctuation $\delta v^2$ of the field (\ref{72}) on
 this stage is expressed for two first moments of $B$ as
 \begin{equation}
 \delta v^2=\frac{\langle B^2\rangle}{\langle B\rangle ^2}-1 \simeq
 \frac{l_0}{L_0}\,\frac 12\,e^{Re^2_0},
 \label{117}
 \end{equation}
 and when the condition (\ref{107}) is valid it is very large in
 contrast to the nonlinear stage (\ref{105}), where $\delta v^2\ll
 1$.
 For the noise carrier with the scale $l_0\ll L_0$ the relative
 integral fluctuation is the small factor $l_0/L_0$ times the
 fluctuation of the simple pulse with random Gaussian amplitude
 (\ref{78}).

 The calculation of $\langle B\rangle ,\langle\triangle B^2\rangle$
 may be done directly on the base of the integral (\ref{b}), and for
 large Reynolds number we have the same equation (\ref{117}).
 For $(l_0/L_0)\ll \exp[-Re_0^2]$ we have relatively small
 fluctuation of $\delta v^2$, and with increasing of $l_0/L_0$ the
 probability distribution of $B$ approaches slowly the normal
 distribution (normalization).
 This normalization is similar to
 the normalization of the homogeneous field at the old-age stage
 \cite{AMS94}.
 But while the moment $B^n$ increases exponentionally
 with $n$, we have weak convergence to the normal distribution of
 $B$ like in \cite{AMS94}.

 \section{Discussion and conclusion}

 We have investigated the evolution of pulses with complex structure in
 nonlinear media, described by Burgers' equation.
 The investigation for vanishing viscosity was done in our paper \cite{GEP99}.
 There it was shown that the asymptotic form of an arbitrary initial
 pulse with zero area is an N-wave.
 It was also shown that the shock positions of the N-wave are determined by
 only  one parameter of the
 initial perturbation - the value of the absolute maximum of the initial
 potential $\psi_{0}(x)$ ($v_{0}(x)=-\partial\psi_{0}(x)/\partial x$).
 It was also shown in the paper \cite{GEP99} that for
 the noise carrier the fluctuations of the shock positions are relatively
 small if the carrier length
 scale $l_{0}$ is much smaller than the modulation length scale $L_{0}$.

 In the present paper we have considered the evolution of pulses with
 complex structure at large but finite Reynolds number.
 On the base of the Hopf-Cole solution it is shown that the finite viscosity
 leads to a finite shock width $\delta$, which does not depend on the fine
 structure of the initial pulse and is fully determined by the shock position
 in the zero viscosity limit.
 The other effect is the shift of the shock position from the zero viscosity
 limit position.
 This shift depends on the fine structure of the initial pulse, and as a
 consequence the time $t_{lin}$, at which the nonlinear stage
 of evolution changes to the linear stage, is determined not only by the
 value of the maximum of the initial potential but also by the fine structure
 of the pulse.
 Because the amplitude of the pulse at
 the linear old-age stage is determined by the time $t_{lin}$, the
 old-age amplitude is also sensitive to the inner structure of the
 pulse.

 In this paper the evolution of a simple N-pulse with regular and random
 initial amplitude and of
 pulses with monochromatic and noise carrier is considered.
 It is shown that the nonlinearity of the
 medium leads to the generation of a non-zero mean field from an initial
 random field with zero mean value.
 It is also shown that, at the nonlinear stage, the relative
 fluctuation of the field and its
 energy is of unit order for simple pulses and small for pulses with complex
 inner structure ($l_{0}<<L_{0}$).
 However, at the old-age linear stage this fluctuation increases exponentially
 with increasing initial Reynolds number.

 \section*{Acknowledgements}

 This work was supported by a grant from the Royal Swedish Academy of
 Sciences, by the grant INTAS No 97-11134  and by the RFBR grant No 99-02-
 18354.
 Sergey Gurbatov thanks the staff of the Department of Mechanics
 (KTH) and other friends in Sweden for their warm hospitality.
 \pagebreak

\begin{figure}[h]
\centerline{\psfig{file=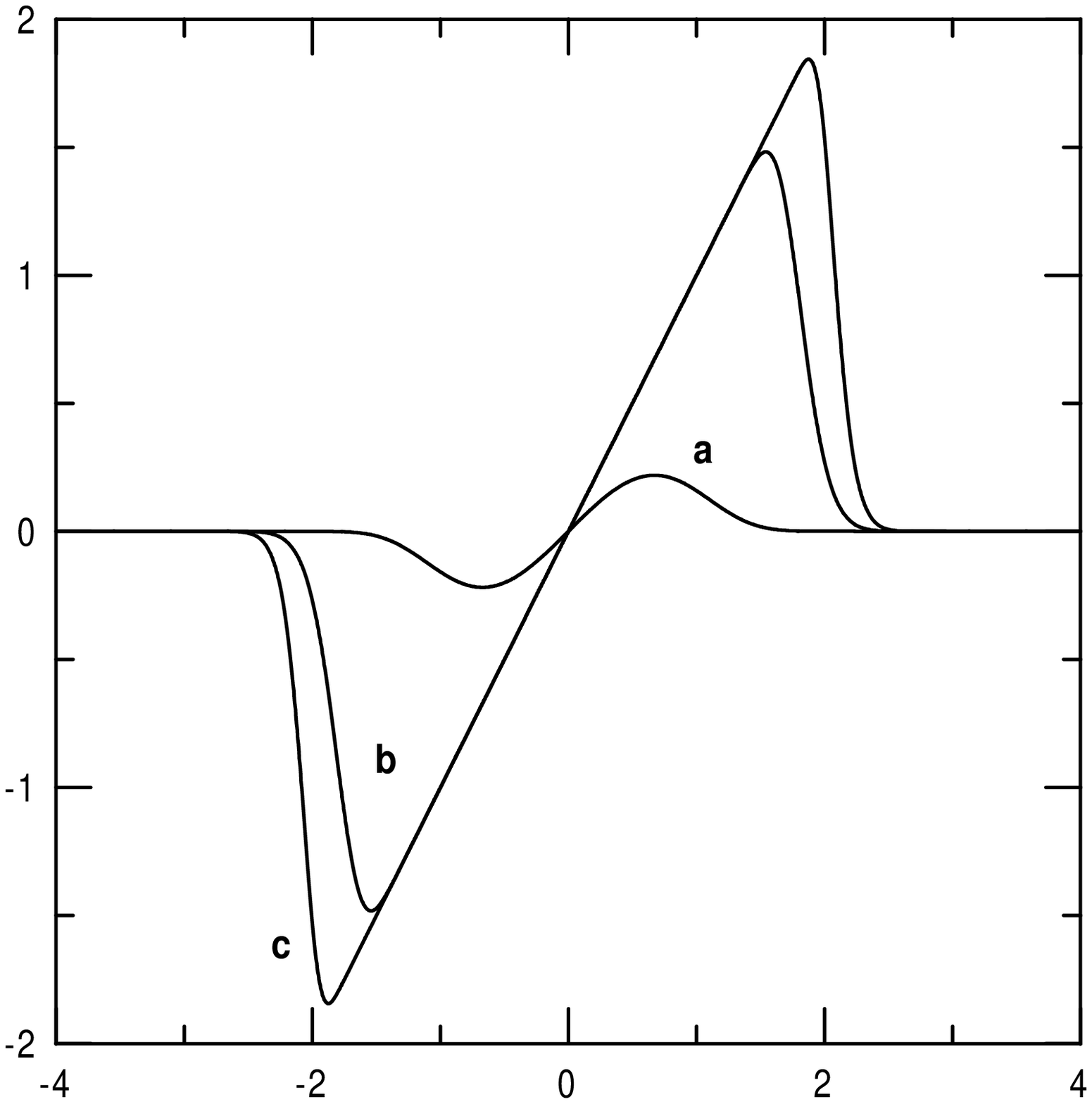,width=10cm,clip=}}
\caption
{Self-similar mean velocity field form $\bar V(y)$ for the
pulse with random initial amplitude (\ref{71}) (curve {\bf a}) and
for the pulse with noise carrier (\ref{98}) for $N=100$ (curve
{\bf b}) and $N=1000$ (curve {\bf c}).
}
\label{f:fig1}
\end{figure}

\clearpage
\pagebreak

\begin{figure}[h]
\centerline{\psfig{file=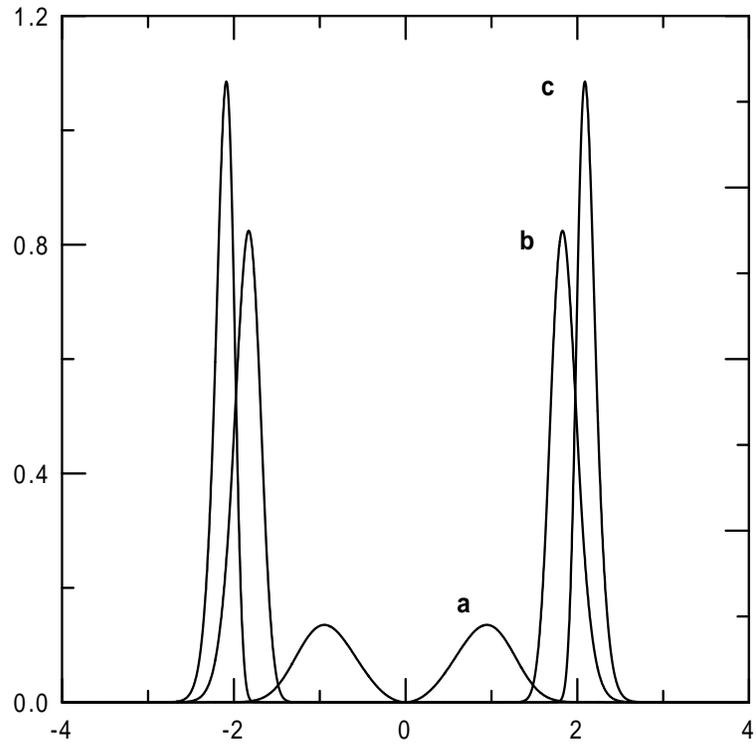,width=10cm,clip=}}
\caption
{Self-similar fluctuation form $\bar\sigma^2_V(y)$ for the
pulse with random initial amplitude (\ref{71}) (curve {\bf a}) and
for the pulse with noise carrier (\ref{98}) for $N=100$ (curve
{\bf b}) and $N=1000$ (curve {\bf c}).}
\label{f:fig2}
\end{figure}

\clearpage
\pagebreak

 \end{document}